\documentclass[aps,pra,superscriptaddress,showpacs,showkeys,a4paper,10pt,notitlepage,twocolumn]{revtex4-1}
\usepackage[utf8]{inputenc}
\usepackage[english]{babel}
\usepackage{amsmath}
\usepackage{amssymb}
\usepackage{amsfonts}
\usepackage{graphicx}
\usepackage{cancel}
\usepackage{upgreek}
\usepackage{mathtools}
\usepackage[T1]{fontenc}

\usepackage{xcolor}
\definecolor{link_blue}{RGB}{52,46,157}

\usepackage{longtable}
\usepackage[colorlinks,citecolor=link_blue,linkcolor=link_blue,urlcolor=link_blue]{hyperref}
\usepackage[tiny,center,uppercase]{titlesec}

\renewcommand{\vec}{\boldsymbol}

\begin{document}

\title{Radiation induced interaction potential of two qubits
  strongly coupled with a quantized electromagnetic field}

\author{I.\ D.\ Feranchuk}
\affiliation{Atomic Molecular and Optical Physics Research Group,
  Advanced Institute of Materials Science, Ton Duc Thang University,
  Ho Chi Minh City, Vietnam}
\affiliation{Faculty of Applied Sciences, Ton Duc Thang University, Ho
  Chi Minh City, Vietnam}

\author{N.\ Q.\ San}
\affiliation{Belarusian State University, 4 Nezavisimosty Ave.,
  220030, Minsk, Belarus}

\author{A.\ U.\ Leonau}
\affiliation{Belarusian State University, 4 Nezavisimosty Ave.,
  220030, Minsk, Belarus}

\author{O.\ D.\ Skoromnik}
\email[Corresponding author: ]{olegskor@gmail.com}
\affiliation{Ho Chi Minh City University of Education, 280 An Duong
  Vuong, District 5, Ho Chi Minh City, Vietnam}

\begin{abstract}
  We investigate the interaction of two two-level qubits with a single
  mode quantum field in a cavity without rotating wave approximation
  and considering that qubits can be located at an arbitrary distance
  from each other. We demonstrate that there exists a radiation induced
  interaction potential between atoms. We studied the properties of
  the system numerically and in addition constructed a simple
  analytical approximation. It is shown that the observable
  characteristics are substantially dependent on the distance between
  the qubits in the strong coupling regime. This allows one to perform
  the quantum control of the qubits, which can be exploited for the
  recording and transmission of quantum information.
\end{abstract}

\maketitle

\section{Introduction}
\label{sec:introduction}

Quantum Rabi model (QRM) describes the interaction of a two-level atom
with a resonant single-mode quantum field in a cavity
\cite{PhysRev.49.324, PhysRev.51.652}. This model plays fundamental
role in the radiation-matter interaction in cavity quantum
electrodynamics \cite{PhysRevA.99.033823, PhysRevA.99.033834,
  PhysRevLett.122.123604}, quantum optics \cite{Walther_2006}, quantum
information \cite{RevModPhys.73.565} and physics of condensed matter
\cite{holstein_studies_1959}. Recently, the model attracted attention
due to the fact that in many systems it is possible
\cite{RevModPhys.91.025005, frisk_kockum_ultrastrong_2019} to control
the interaction strength in a wide range, including the so called
ultra strong coupling (USC) regime, which corresponds to the variation
interval from 0.3 to 1.0 of a dimensionless coupling constant $f$
between an atom (qubit) and a field. The systems with the coupling
constant from the USC range were lately realized experimentally
\cite{PhysRevA.96.013849, RevModPhys.91.025005}. As a result, these
achievements are crucial for control of an interaction of quantum
emitters with individual photons that is an important part of
recording and transmission of quantum information.

Another related direction is the generalization of the QRM to systems
containing multiple qubits, in particular the two qubits interacting
with a resonance quantum field - the Tavis-Cummings model (TCM)
\cite{PhysRev.170.379, PhysRev.188.692}. This model is based on two
approximations. The first is the rotating wave approximation (RWA)
applicable for small values of the detuning between frequencies of the
field and the resonant atomic transition and small values of the
coupling constant $f$ of the atom-field interaction. The second
approximation assumes that the distance
$\rho = |\vec{R}_{1} - \vec{R}_{2}|$ between the qubits is small in
comparison with the wavelength $\lambda$ of the resonance field, i.e.,
$\rho \ll \lambda$. In addition, in works \cite{Agarwal_2012,
  PhysRevA.99.033834} the TCM was investigated beyond the RWA,
however, still under the assumption $\rho \ll \lambda$.

At the same time, the systems containing two qubits located at the
distance $\rho \sim \lambda$ and interacting with a resonant quantum
field were recently realized experimentally. Moreover, it was possible
to control the positions of qubits in a broad range with the use of
tightly focused optical tweezers \cite{Twoqubit1, Twoqubit2}. Similar
experiments were recently performed for systems of Rydberg atoms
located inside a cavity \cite{Twoatomscavity,Twoatomstrapping}. As a
result, in our work we theoretically investigate the spectrum and
dynamics of the system in the USC regime beyond the RWA and as a
function of the distance $\rho$ between qubits. We demonstrate that
the dependence on $\rho$ becomes important in the USC regime where the
RWA in not applicable.

In our work we employ a dipole approximation for an individual qubit
interacting with a quantum field. This approximation is consistent
with the assumption that $\rho \sim \lambda$ due to the fact that for
optical frequencies of the resonance field the distance $\rho$ between
qubits is much larger than the characteristic qubit size $a_{0}$,
therefore $\rho \gg a_{0}$ allowing us to work in the dipole
approximation for each qubit. In addition, the large qubit mass allows
us to employ the adiabatic approximation in an analogy with the
Born-Oppenheimer approximation of molecular physics. As a result, we
treat the operator of kinetic energy of qubits as a perturbation. We
show that the energy levels of the systems form potential surfaces as
a function of the distance $\rho$, which define the radiation induced
potential of the interaction between qubits. The form of this
potential is defined by the coupling constant $f$.  Moreover, the
distance between qubits can be considered as an additional parameter
to be used to control the system's characteristics for the recording
and the transmission of quantum information.

In addition, the dependence of observable characteristics of the
system of two qubits on the distance between them adds an additional
possibility to control the system. In particular, this allows one to
control the location of the peak of in the scattering cross section of
the resonance radiation \cite{Scattering_2020}; to vary the degree of
entanglement when the transmission of quantum information is happening
by two emitters (qubits) \cite{PhysRevA.101.032335}; to change the
population of states of two two-level systems to control the
probability of a spontaneous emission
\cite{feranchuk_spontaneous_2017}; to obtain a periodic structure in
the system of $N$-atoms --- the extended Dicke model
\cite{PhysRevA.99.013839}.

\section{Construction of a model Hamiltonian}
\label{sec:constr-model-hamilt}

The Hamiltonian of two identical two-level atoms (qubits) with the
mass $M$, located at positions $\vec{R}_{1}$ and $\vec{R}_{2}$ in the
dipole approximation for the interaction of atoms with the field,
written in natural units ($\hbar = c = 1$) reads
\cite{scully_quantum_1997, di_stefano_resolution_2019}
\begin{align}
  \hat H
  &= - \frac{1}{2M}\left(\Delta_{\vec R_1} + \Delta_{\vec R_2}\right)
    + \frac{\epsilon}{2} \left(\sigma^1_3 + \sigma^2_3\right) \nonumber
  \\
  & + \omega f \Big[\left(\hat{a} e^{i \vec k \cdot \vec R_1} +
    \hat{a}^{\dag} e^{-i \vec k \cdot \vec R_1} \right) \sigma^1_1 \nonumber
  \\
  & \mspace{60mu}+ \left(\hat{a} e^{i \vec k \cdot \vec R_2} +
    \hat{a}^{\dag} e^{-i\vec k \cdot \vec R_2} \right)\sigma^2_1\Big]
    \nonumber
  \\
  &+ \omega \hat{a}^{\dag}\hat{a} + V_a (\vec R_1 - \vec
    R_2), \label{eq:1}
  \\
  f
  &=  e_0\omega\Delta d\sqrt{\frac{4\pi }{\omega^3V}}. \label{eq:2}
\end{align}
Here $V$ is the volume of the cavity, $f$ is the dimensionless
coupling constant of an atom-field interaction, $\epsilon$ is the
resonance transition energy between two qubit states
$\chi_{\uparrow}$, $\chi_{\downarrow}$ with the dipole transition
matrix element $d$; $e_{0}$, $m_{0}$ are the electron charge and mass
respectively; $\hat{a}^{\dag}$, $\hat{a}$ are the creation and
annihilation operators of the resonant quantum field with the
frequency $\omega$ and the wave vector $\vec k$,
$V_{a}(\vec{R}_{1} - \vec{R}_{2})$ is the atom-atom interaction
potential due to the exchange and dipole-dipole interactions, $\Delta$
is the Laplace operator and $\vec{\sigma}^{1}$, $\vec{\sigma}^{2}$ are
Pauli matrices for qubits one and two respectively. The limit of
$M \to \infty$ corresponds to the situation of two immovable qubits.

Let us switch to the center of mass reference system in Eq.~(\ref{eq:1})
\begin{equation}
  \vec R = \frac{\vec R_1 + \vec R_2}{2}, \quad
  \vec \rho = \vec R_1 - \vec R_2, \label{eq:3}
\end{equation}
in which the Hamiltonian (\ref{eq:1}) transforms into
\begin{align}
  \hat H
  &= - \frac{1}{4M}\Delta_{\vec R}  - \frac{1}{M} \Delta_{\vec \rho }
    + \frac{\epsilon}{2} (\sigma^1_3 + \sigma^2_3) \nonumber
  \\
  & + \omega f \Bigg[\left(\hat{a} e^{i \vec k \cdot(\vec R + \vec \rho/2)}
    + \hat{a}^{\dag} e^{- i \vec k \cdot (\vec R + \vec \rho/2)}\right)
    \sigma^1_1 \nonumber
  \\
  &\mspace{60mu}+ (\hat{a} e^{i \vec k \cdot (\vec R - \vec \rho/2)}
    + \hat{a}^{\dag} e^{-i \vec k \cdot (\vec R - \vec
    \rho/2)})\sigma^2_1 \Bigg] \nonumber
  \\
  &+ \omega \hat{a}^{\dag} \hat{a} + V_a (\vec \rho). \label{eq:4}
\end{align}

The system possesses translational invariance with respect to the
center of mass coordinate $\vec{R}$. Therefore, in analogy with the
polaron problem we perform the Lee-Low-Pines transformation \cite{LLP}
\begin{align}
  \hat H' = \hat L^{-1}\hat H \hat L, \quad
  \hat L = e^{i (\vec P - \vec k \hat{a}^{\dag}\hat{a}) \cdot \vec
  R}, \label{eq:5}
\end{align}
where $\vec P$ is the total momentum of the system, which in this case
is an integral of motion. As a result, we find the following expression
for the Hamiltonian of the system
\begin{align}
  \hat H'
  &=  \frac{1}{4M}(\vec P - \vec k \hat{a}^{\dag} \hat{a})^2
    - \frac{1}{M}\Delta_{\vec \rho } + \frac{\epsilon}{2}
    (\sigma^1_3 + \sigma^2_3) \nonumber
  \\
  &+\omega f \Bigg[\left(\hat{a} e^{i \vec k \cdot \vec \rho / 2} +
    \hat{a}^{\dag} e^{-i \vec k \cdot \vec \rho/2}\right) \sigma^1_1
    \nonumber
  \\
  &\mspace{60mu} + (\hat{a} e^{- i \vec k \cdot \vec \rho/2 } +
    \hat{a}^{\dag} e^{ i \vec k \cdot \vec \rho/2 })\sigma^2_1 \Bigg]
    \nonumber
  \\
  &+ \omega \hat{a}^{\dag}\hat{a} + V_a (\vec \rho). \label{eq:6}
\end{align}

The characteristic scale with respect to the coordinate of a relative
motion $\vec{\rho}$ is defined by the wavelength of the radiation
$\lambda = 2\pi / k$ and for the optical frequencies substantially
exceeds the size of the atom $a_{0}$, such that
$a_{0} / \lambda \sim 10^{-3}$. At these distances the contribution
from the exchange interaction into the potential $V_{a}(\vec \rho)$ is
exponentially suppressed. The dipole-dipole interaction potential (the
Van der Walls) of an atom-atom interaction is proportional
\cite{landau_quantum_2007} to
$\kappa(a_{0}/\rho)^{6} \sim \kappa(a_{0}/\lambda)^{6}$, with $a_{0}$
being the characteristic atomic size and $\kappa$ is the dimensional
constant. Consequently, in the dipole approximation
($a_{0} / \lambda \ll 1$) for the interaction of an individual atom
with the electromagnetic field, the potential $V_{a}(\vec\rho)$
should be neglected in the Hamiltonian $\hat{H}'$ (\ref{eq:6}) for the
consistency. In addition, we can also neglect a recoil in the operator
of kinetic energy since we are working in the nonrelativistic
limit. Consequently, one can write
\begin{align}
  \langle \frac{1}{2M}(\vec P  \vec k \hat{a}^{\dag} \hat{a}) \rangle
  \approx \frac{P}{2M}\omega \ll \omega. \label{eq:7}
\end{align}

\begin{figure*}[t]
  \includegraphics[width=\columnwidth]{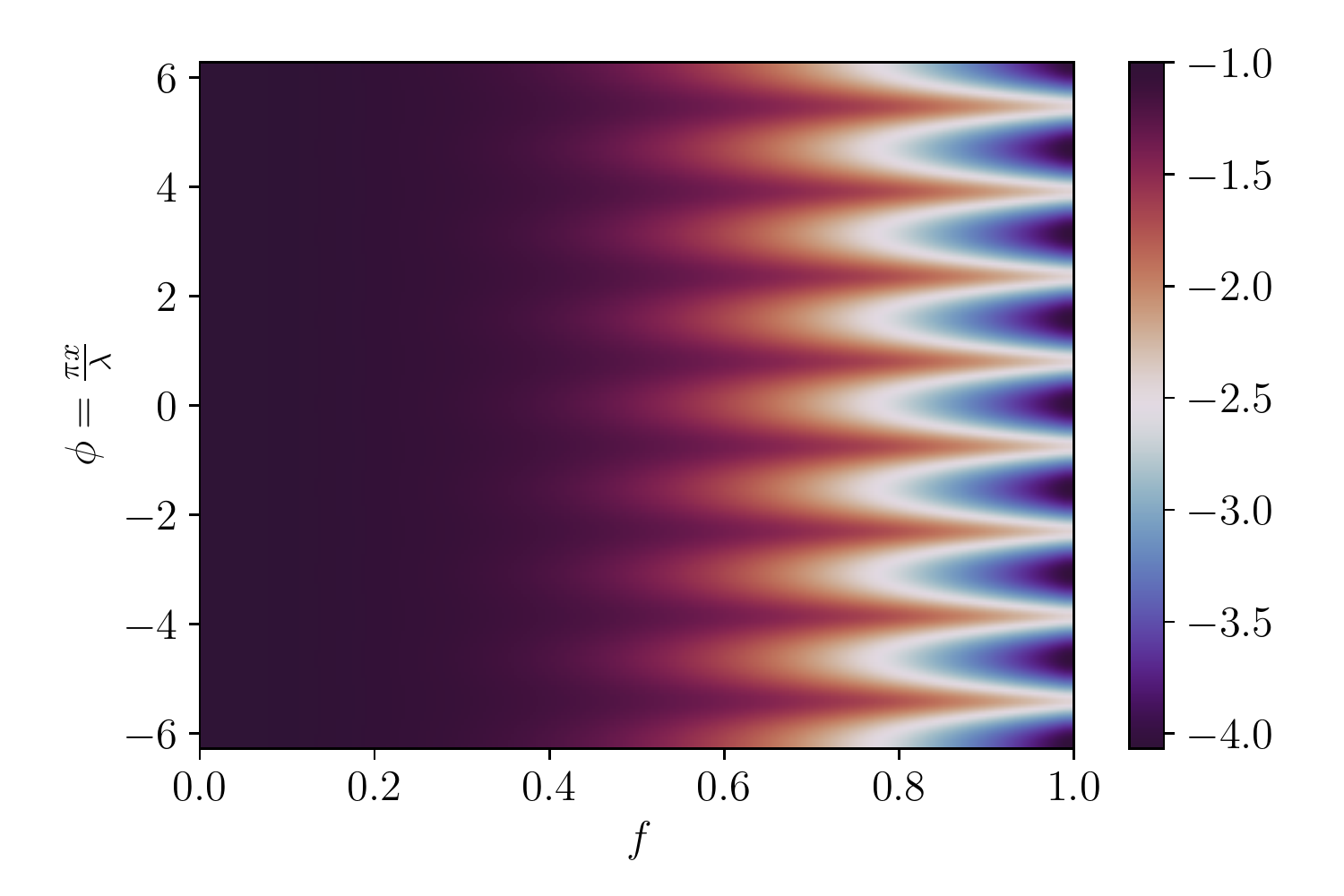}
  \includegraphics[width=\columnwidth]{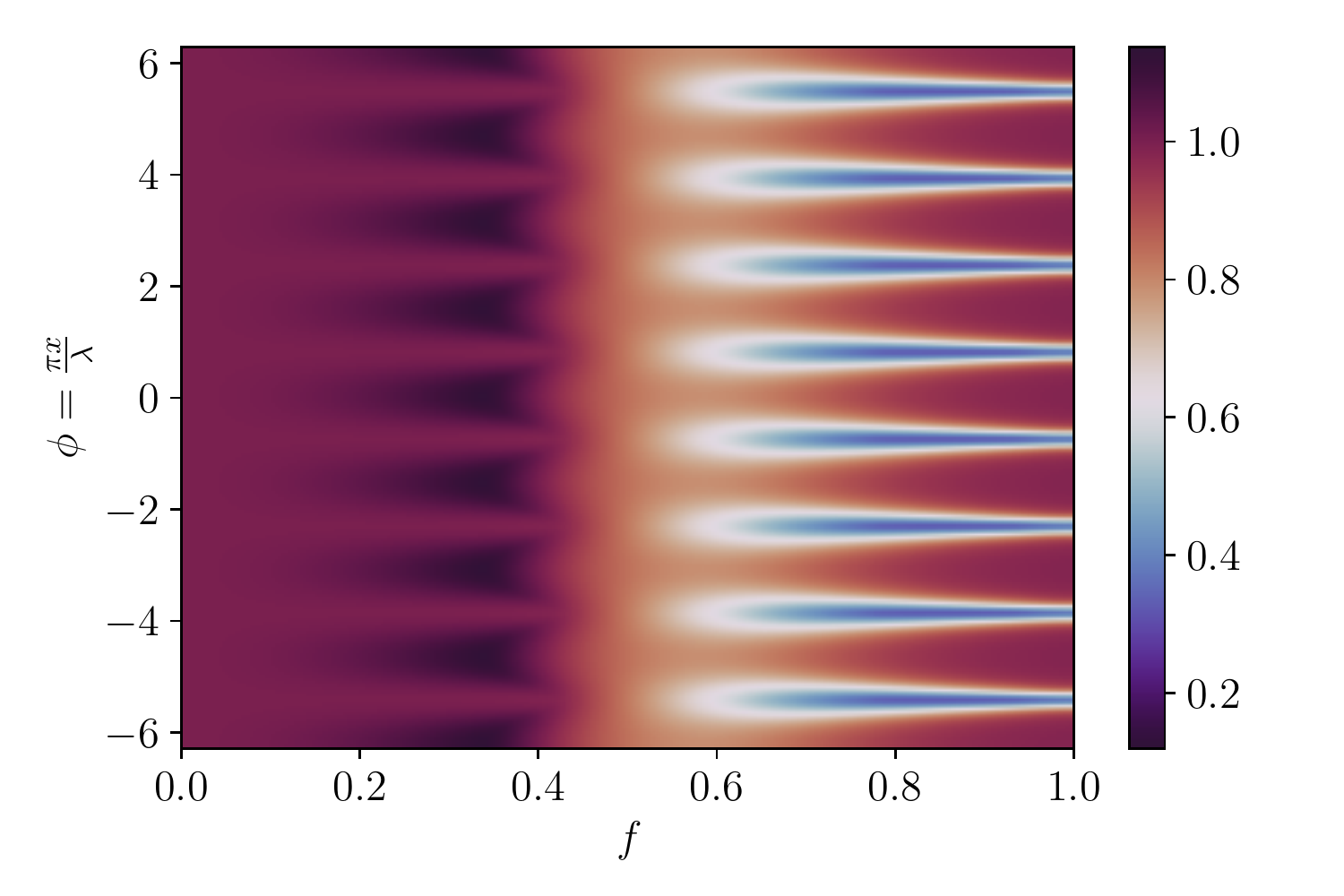}
  \caption{(Color online) Left panel. The potential surface
    $u_{0}(\phi)$ of the ground state as a function of the
    dimensionless coupling constant $f$ and the coordinate
    $\phi = \pi x / \lambda$. The parameter $\delta = 1$. When the
    system of two qubits is in the USC regime the potential wells are
    the deepest. Right panel. The transition frequency
    $\Omega_{\nu_{1},\nu_{2}}(\phi)$ from the ground state to the
    third excited state as a function of the dimensionless coupling
    constant $f$ and the coordinate $\phi$. The parameter
    $\delta = 1$.}\label{fig:1}
\end{figure*}

As a result, we arrive to the final expression for the Hamiltonian,
which describes the interaction of two two-level atoms within the
above described approximations
\begin{align}
  \hat H'
  &=  \frac{P^{2}}{4M} - \frac{1}{M}\Delta_{\vec \rho }
    + \frac{\epsilon}{2} (\sigma^1_3 + \sigma^2_3) + \omega
    \hat{a}^{\dag}\hat{a} \nonumber
  \\
  &+\omega f \Bigg[\left(\hat{a} e^{i \vec k \cdot \vec \rho / 2} +
    \hat{a}^{\dag} e^{-i \vec k \cdot \vec \rho/2}\right) \sigma^1_1
    \nonumber
  \\
  &\mspace{60mu} + (\hat{a} e^{- i \vec k \cdot \vec \rho/2 } +
    \hat{a}^{\dag} e^{ i \vec k \cdot \vec \rho/2 })\sigma^2_1 \Bigg].
    \label{eq:8}
\end{align}

\section{The interaction potential of two qubits}
\label{sec:inter-potent-two}

Let us investigate the Schr\"{o}dinger equation with the Hamiltonian
(\ref{eq:8}). For this purpose we choose a coordinate system in which
the $x$-axis is directed along the $\vec{k}$ and seek a solution in
the form $\Lambda(\vec{\rho}) = \exp\{-i(p_{y}y + p_{z}z)\}\Psi(x)$
\begin{align}
  \Bigg\{
  &\frac{P^{2}}{4M} + \frac{(p_{y}^{2} + p_{z}^{2})}{M} -
    \frac{1}{M}\frac{d^{2}}{dx^{2}}
    + \frac{\epsilon}{2} (\sigma^1_3 + \sigma^2_3) + \omega
    \hat{a}^{\dag}\hat{a} \nonumber
  \\
  &+\omega f \Bigg[\left(\hat{a} e^{i \phi} +
    \hat{a}^{\dag} e^{-i \phi}\right)
    \sigma^1_1 \nonumber
  \\
  &\mspace{60mu} + (\hat{a} e^{- i \phi} +
    \hat{a}^{\dag} e^{i\phi})\sigma^2_1
    \Bigg]\Bigg\}\Psi(x) = E \Psi(x), \label{eq:9}
\end{align}
where $\phi = \pi x / \lambda$.

The case of $x = 0$ corresponds to the situation that atoms are
unified and form a system with different parameters. Therefore, we
assume that in the operator (\ref{eq:9}) the coordinate $x$ is varying
in the range $|x| > x_{0}$, where the quantity
$x_{0} \sim a \ll \lambda$. In the general case the vector
$\vec{\rho}$ can be directed under an arbitrary angle with respect to
the vector $\vec k$. However, we consider that the following
conditions $y \approx z \approx x_{0}$ are fulfilled for the
projections of $\vec\rho$ on the direction perpendicular to the $\vec{k}$
vector.

As it was explained in the introduction, the operator of kinetic
energy of a qubit is a small correction in comparison with the
operator of the interaction of a qubit and a resonance
field. Therefore, we employ an adiabatic approximation and separate
variables in Eq.~(\ref{eq:9})
\begin{align}
  &\Psi(x)\approx \Phi(x) \psi_{\nu}, \label{eq:10}
  \\
  &\Bigg\{\frac{\epsilon}{2} (\sigma^1_3 + \sigma^2_3) + \omega
    \hat{a}^{\dag}\hat{a} + \omega f \Bigg[\left(\hat{a} e^{i \phi} +
    \hat{a}^{\dag} e^{-i \phi}\right) \sigma^1_1 \nonumber
  \\
  &\mspace{40mu} + (\hat{a} e^{- i \phi} +
    \hat{a}^{\dag} e^{i\phi})\sigma^2_1
    \Bigg]\Bigg\}\psi_{\nu} = U_{\nu}(x)\psi_{\nu}, \label{eq:11}
  \\
  &\Bigg\{\frac{P^{2}}{4M} + \frac{(p_{y}^{2} + p_{z}^{2})}{M} -
    \frac{1}{M}\frac{d^{2}}{dx^{2}} + U_{\nu}(x)\Bigg\} \Phi(x)
    \nonumber
  \\
  &\mspace{45mu}= E \Phi(x). \label{eq:12}
\end{align}
In these equations the index $\nu$ denotes a set of quantum numbers of
qubit-photon system and the terms $U_{\nu}(x)$ define the radiation
induced interaction potential of two qubits in a full analogy with the
molecular physics. The solution of the Schr\"{o}dinger equation
(\ref{eq:12}) determines the relative motion of qubits induced by the
potential $U_{\nu}(x)$.

We want to mention here that when the $\phi$ equals zero the operator
in Eq.~(\ref{eq:11}) coincides with the one obtained in works
\cite{Agarwal_2012, PhysRevA.99.033834}. In this case, $U_{\nu}(0)$
corresponds with the spectrum found in \cite{Agarwal_2012,
  PhysRevA.99.033834}.

In order to calculate the potential function $U_{\nu}(x)$ approximate
methods can be employed \cite{Twoqubit2}. In our work we have
developed the analytical approximation for energy levels and in
addition performed an exact numerical solution by diagonalizing the
Hamiltonian matrix written in the basis of eigenstates of $\sigma_{3}$
and Fock states of the field. The details of the calculations are
given in appendix \ref{appendix:A}.

From the structure of Eq.~(\ref{eq:11}) it follows that the
Hamiltonian of the system is a periodic function with a period
$2\lambda$.

It is convenient to express the energies in the units of the photon
frequency
\begin{align}
  U_{\nu}(x)
  &= \omega u_{\nu}(x), \label{eq:13}
  \\
  \Bigg\{\frac{\delta}{2} (\sigma^1_3
  &+ \sigma^2_3) + \hat{a}^{\dag}\hat{a} + f \Bigg[\left(\hat{a} e^{i
    \phi} + \hat{a}^{\dag} e^{-i \phi}\right) \sigma^1_1 \nonumber
  \\
  & + (\hat{a} e^{- i \phi} +
    \hat{a}^{\dag} e^{i\phi})\sigma^2_1
    \Bigg]\Bigg\}\psi_{\nu} = u_{\nu}(x)\psi_{\nu}, \label{eq:14}
\end{align}
where we introduced $\delta = \epsilon / \omega$.

We plot in Fig.~\ref{fig:1} (Left panel) the potential surface as a
function of the coupling constant $f$ for the ground state of the
system when $\delta = 1$. The depth of the potential wells is varying
by two orders of magnitude when the coupling constant is changing in
the range $\in [0,1]$. When the system is in the USC regime and
approaching $f = 1$ the depth of the wells is four times larger (i.e.
$4\delta$) than the splitting $\delta$ between qubit energy levels. As
a result, we will be mainly investigating this most interesting
regime.

We pay attention to the fact that the period of oscillations of the
potential equal to $\lambda/2$, which is different from the
periodicity $2\lambda$ of the Hamiltonian of the system. This symmetry
arises also in the analytical approximation derived in the appendix
\ref{appendix:A} (See also Fig.~\ref{fig:5}). The decrease of the
period of the oscillations of the interaction potential is explained
by the fact that the eigenvalues of the Hamiltonian in
Eq.~(\ref{eq:14}) are degenerate --- the eigenvalues are invariant
under the transformation $R_j^{-1} H R_j$ of the Hamiltonian,
$j = 1, 2, 3$ with
$R_{1,2} = \exp\{\pm i \pi \hat a^\dag \hat a / 2\}$ and
$R_3 = \exp\{i \pi \hat a^\dag \hat a\}$.

We also note here, that the dependence of the energy of the ground
state on $x$ is caused mainly by the counter rotating terms in the
Hamiltonian (\ref{eq:8}). This follows from the fact that in the RWA
the exact ground state of the system is given by
$\psi_{0}^{\mathrm{RWA}} =
\chi^{1}_{\downarrow}\chi^{2}_{\downarrow}|0\rangle$ and by acting
with $\hat{H}'$ on $\psi_{0}^{\mathrm{RWA}}$ one finds that
$u_{\nu}^{\mathrm{RWA}}(x) = -\delta$ and is independent of $x$. Here
$\chi^1_{\downarrow}, \chi^2_{\downarrow}$ are the ground states of
the first and second qubits respectively and $|0\rangle$ is the vacuum
state of the field.

\section{Exact numerical solution}
\label{sec:exact-numerical-solution}

Let us present how we performed the exact numerical solution for the
analysis of the system's dynamics. For this, we consider that both
atoms in the initial moment of time occupy the down states
$\chi_\downarrow^1$ and $\chi_\downarrow^2$ and the field is prepared
in the coherent state with the amplitude $\alpha = \sqrt{\bar{n}}$
($\bar{n}$ is the average number of photons)
\begin{equation}
  |\Psi (0)\rangle = |\alpha\rangle \chi_\downarrow^1
  \chi_\downarrow^2. \label{eq:15}
\end{equation}

The functions $\chi_\downarrow^{1,2}$ and $\chi_\uparrow^{1,2}$ are
the eigenfunctions of $\sigma_{3}$. We also use the following
convention and ordering for the extended spin space unit base vectors:
\begin{align}
  |\chi_1\rangle \equiv \chi_\uparrow^1 \chi_\uparrow^2 =
  \begin{pmatrix}
    1 \\ 0 \\ 0 \\ 0
  \end{pmatrix}, \quad
  |\chi_2\rangle \equiv \chi_\uparrow^1 \chi_\downarrow^2 =
  \begin{pmatrix}
    0 \\ 1 \\ 0 \\ 0
  \end{pmatrix}, \nonumber \\
  |\chi_3\rangle \equiv \chi_\downarrow^1 \chi_\uparrow^2 =
  \begin{pmatrix}
    0 \\ 0 \\ 1 \\ 0
  \end{pmatrix}, \quad
  |\chi_4\rangle \equiv \chi_\downarrow^1 \chi_\downarrow^2 =
  \begin{pmatrix}
    0 \\ 0 \\ 0 \\ 1
  \end{pmatrix} \label{eq:16}
\end{align}
First one needs to obtain the exact numerical solution of the
eigenvalue problem. For this purpose we introduce the following matrix
elements of the Hamiltonian (\ref{eq:14}) in the Fock field states and
spin base vectors (\ref{eq:16}):
\begin{align}
  H_{nk}
  &= \left( \frac{\delta}{2} (\sigma_3 \otimes \text{I}_2 +
    \text{I}_2 \otimes\sigma_3) + n \text{I}_4 \right)
    \delta_{nk} \label{eq:17}
  \\
  &+ f \Bigl[ (\sqrt{k} e^{i \phi} \delta_{n,k-1} + \sqrt{k+1} e^{- i
    \phi} \delta_{n,k+1}) \sigma_1 \otimes \text{I}_2 \nonumber
  \\
  &+ (\sqrt{k} e^{-i \phi} \delta_{n,k-1} + \sqrt{k+1} e^{ i \phi}
    \delta_{n,k+1}) \text{I}_2 \otimes\sigma_1  \Bigr], \nonumber
\end{align}
where $\text{I}_4$ is a unit $4\times4$ matrix, $\text{I}_2$ is the
unit $2\times2$ matrix.

By numerically solving the eigenvalue problem for (\ref{eq:17}), we
obtain the set of eigenvalues $\{E_{\varkappa}\}$ and corresponding
eigenvectors $\{|\psi_{\varkappa}\rangle\}$ (in the form of a list of
expansion coefficients $\{C^\varkappa_{kq}\}$ for each
eigenvector). As a result, one can construct the time-dependent wave
vector as an expansion over the stationary states:
\begin{equation}
  |\Psi (t)\rangle = \sum_{\varkappa} A_{\varkappa}
  |\psi_{\varkappa}\rangle e^{-i E_{\varkappa} t}, \label{eq:18}
\end{equation}
where
\begin{equation}
  |\psi_{\varkappa}\rangle = \sum_{k=0}^{\infty} \sum_{q=1}^{4}
  C^{\varkappa}_{nq} |k\rangle |\chi_q\rangle = \sum_{k=0}^{\infty}
  \begin{pmatrix}
    C^{\varkappa}_{k1}\\
    C^{\varkappa}_{k2}\\
    C^{\varkappa}_{k3}\\
    C^{\varkappa}_{k4}
  \end{pmatrix} |k\rangle \label{eq:19}
\end{equation}
The coefficients $A_{\varkappa}$ can be obtained from the initial
condition:
\begin{equation}
  A_{\varkappa} = \langle \psi_{\varkappa}|\Psi(0)\rangle =
  \sum_{k=0}^{\infty} \left(C^{\varkappa}_{k4}\right)^*
  \frac{\alpha^k}{\sqrt{k!}} e^{-\alpha^2 / 2} \label{eq:20}
\end{equation}
The density matrix of the system is
\begin{equation}
    \hat{\rho} = |\Psi (t)\rangle \langle\Psi (t)|
    \label{eq:21}
\end{equation}
and the density matrix of the atomic subsystem can be calculated by
tracing out the field degrees of freedom in Eq.~(\ref{eq:21})
\begin{equation}
  \hat{\rho}_{\mathrm{TQ}} = \sum_{n=0}^{\infty} \langle n|\Psi(t)\rangle
  \langle \Psi(t)|n\rangle \label{eq:22}
\end{equation}
Finally, the probability of finding both atoms in the down state can
be obtained in the way as follows:
\begin{align}
  P_{-1} (t)
  &= \langle \chi_4|\hat{\rho}_{\mathrm{TQ}}|\chi_4\rangle = \nonumber
  \\
  &= \sum_n \left| \sum_{\varkappa} A_{\varkappa} e^{-i E_{\varkappa}
    t} C^{\varkappa}_{n4} \right|^2
  \label{eq:23}
\end{align}

\section{Observable characteristics of the system}
\label{sec:observ-char-syst}

One of the observable consequences of the dependence of the energy of
the system on the distance between qubits is the variation of the
transition frequencies with the variation of $x$, i.e.
\begin{align}
  \Omega_{\nu_{1}\nu_{2}}(x) = \omega(u_{\nu_{2}}(x) -
  u_{\nu_{1}}(x)). \label{eq:24}
\end{align}

\begin{figure*}[t]
  \includegraphics[width=\columnwidth]{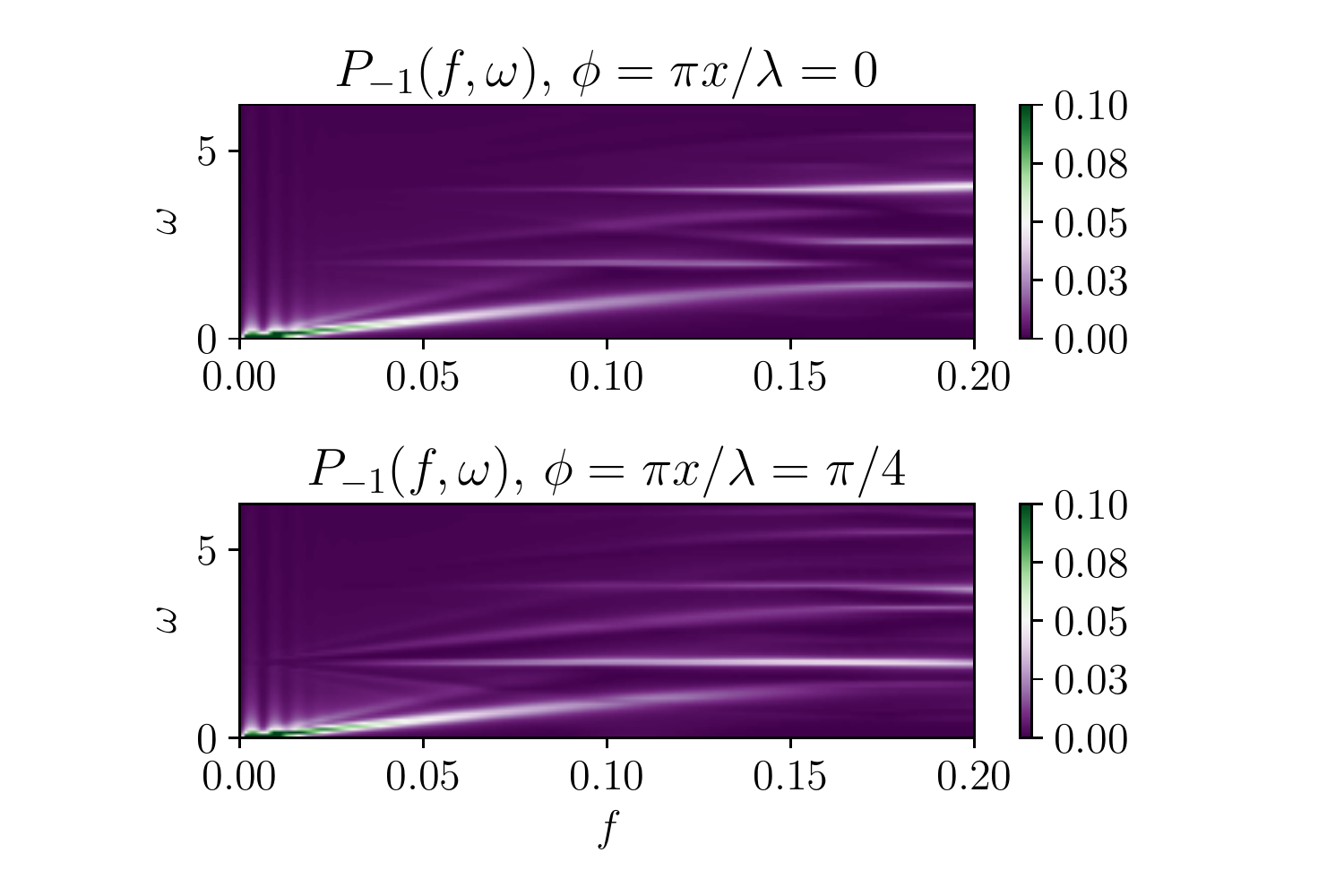}
  \includegraphics[width=\columnwidth]{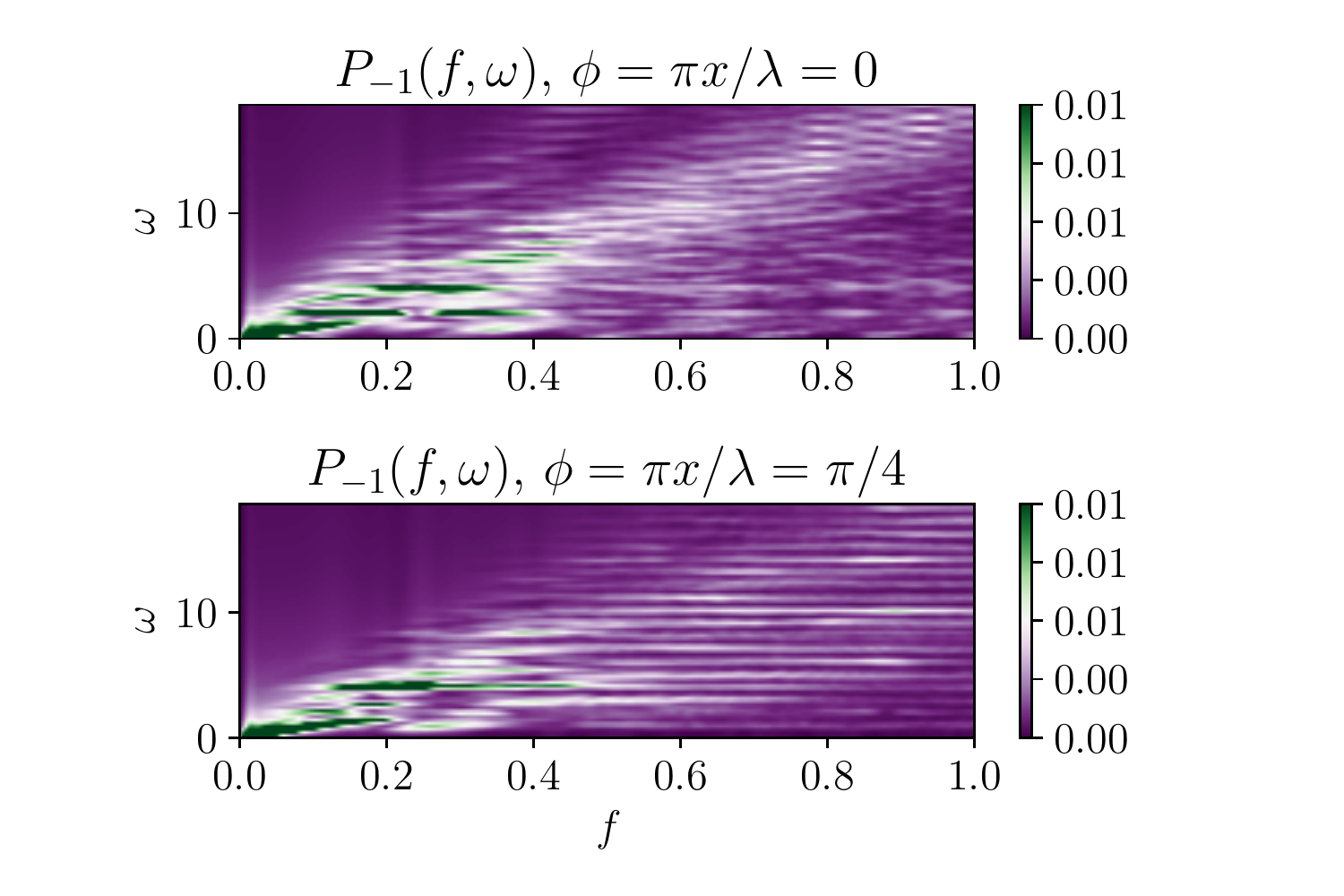}
  \caption{(Color online) Left panel. Fourier transform
    $P_{-1}(f,\omega)$ of the evolution of the population inversion as
    a function of the dimensionless coupling constant $f$, plotted for
    two values of the distance between qubits, $\phi = 0$ and
    $\phi = \pi/4$. The parameter $\delta = 1$ and the field is
    prepared in the coherent state with $\bar n = 25$. When the
    coupling constant increases we observe the doubling of
    frequencies, which signals the transition to the chaotic behavior
    \cite{richtmyer_principles_1981}. Right panel. Fourier transform
    $P_{-1}(f,\omega)$ of the evolution of the population inversion as
    a function of the dimensionless coupling constant $f$, plotted for
    two values of the distance between qubits, $\phi = 0$ and
    $\phi = \pi/4$. The parameter $\delta = 1$ and the field is
    prepared in the coherent state with $\bar n = 25$. When the system
    transitions into the USC regime we observe chaotic behavior, when
    all frequencies appear in the spectrum.}\label{fig:2}
\end{figure*}

In Fig.~\ref{fig:1} (Right panel) we plot this example. The dependence
of the transition frequency on the distance between qubits allows one
to control the scattering cross section of the resonance radiation on
the system \cite{Scattering_2020}. At the same time the change of the
population of quantum states of QRM with $x$ (see Fig.~{\ref{fig:2}})
extends the possibility to control the transition probability and the
linewidths of atomic transitions
\cite{feranchuk_spontaneous_2017}.

Another consequence of the dependence of the transition frequency on
the distance is that for different $x$ the period of the Rabi
oscillations and the time evolution of the system is modified that is
demonstrated in Fig.~\ref{fig:2}. In this Fig. instead of the time
evolution of the ground state in the time domain $P_{-1}(t)$ we plot
its Fourier transform $P_{-1}(\omega)$, which is an even function of
the frequency for the case when the field is prepared in the coherent
state with the average number of photons $\bar n = 25$. As a result,
we plotted only the region of positive frequencies $\omega$. This
quantity demonstrates the changes in the dynamics of the system with
the change of the dimensionless coupling constant $f$. As such, for
small values of the coupling constant (see Fig.~\ref{fig:2} Left
panel) the spectrum possesses a strong maximum located at the
frequency $\omega \approx 5 \sqrt{\bar n} = 5f$, to which coincides a
period of oscillations with only a single Rabi frequency. The
contributions from other frequencies are highly suppressed. When the
coupling constant increases we start to observe the increase of the
amplitudes of other harmonics that are located on different
frequencies. This signifies that the system transitions in the chaotic
regime \cite{richtmyer_principles_1981}. The spectrum is an even
function of the frequency. Consequently, the appearance of the new
maximum brings two new frequencies of an opposite sign (doubling). The
amplitudes of new peaks in the spectrum increase in the region where
the rotating wave approximation is not applicable. This region is
defined by the dimensionless coupling constant
$f \sim 1/\sqrt{\bar{n}}$ (see \cite{leonau_uniformly_2020}). This
behavior of the spectrum can be interpreted as bifurcations. As a
result, in the strong coupling regime (Fig.~\ref{fig:2} Right panel)
the system posseses a spectrum with a broad range of
frequencies. However, we pay attention to the fact that the spectrum
changes with the distance between qubits.

In addition, to the spectrum and dynamics of the system we also
investigate the correlation properties such as the average number of
photons inside the cavity and the entanglement of the qubit
states. The dependence of these quantities on the distance $x$ between
qubits allows one to perform their control. For this purpose let us
investigate these quantities for the ground state of the system.

\begin{figure}[t]
  \centering
  \includegraphics[width=\columnwidth]{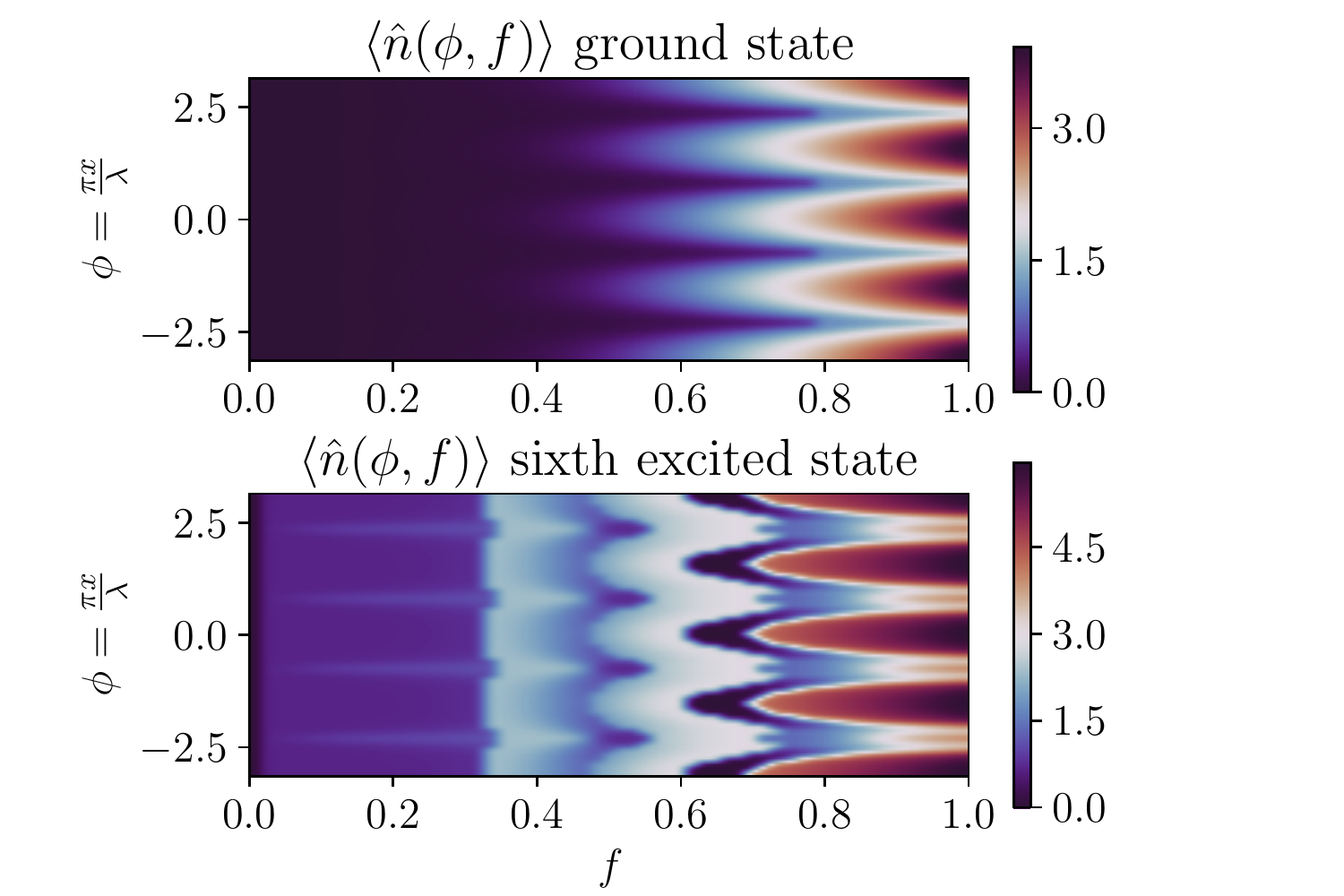}
  \caption{(Color online) The surface of the expectation value of the
    photon number operator
    $\langle n(x,f)\rangle = \langle\hat{a}^{\dag}\hat{a}\rangle$ of
    the ground state and the sixth excited state as a function of the
    dimensionless coupling constant $f$ and the coordinate $\phi$. The
    parameter $\delta = 1$. When the system of two qubits is in the
    USC regime the expectation value is significantly different from
    zero.}\label{fig:3}
\end{figure}

We represent the wave function of the system in the basis of Fock
states $|n\rangle$ and $\chi^{1}_{s_{1}}$, $\chi^{2}_{s_{2}}$ the spin
states of qubits
\begin{align}
  \psi_{0} = \sum_{n, s_{1}, s_{2}}  C^{0}_{n,s_{1},s_{2}}|n\rangle
  \chi^{1}_{s_{1}} \chi^{2}_{s_{2}}, \label{eq:25}
\end{align}
The coefficients $C^{0}_{n,s_{1},s_{2}}$ are computed numerically by
diagonalizing the Hamiltonian matrix in this basis. Consequently, with
the help of this expansion we can compute the average number of
photons inside the cavity as
\begin{align}
  \langle n(x,f)\rangle = \langle\hat{a}^{\dag}\hat{a}\rangle =
  \sum_{n,s_{1},s_{2}} n |C^{0}_{n,s_{1},s_{2}}|^{2}, \label{eq:26}
\end{align}
which we plot in Fig.~\ref{fig:3}. This figure demonstrates that in
the USC regime even when the system is in the ground state the
interaction between qubits and the quantum field leads to the
excitation of three photons. However, this number significantly
depends on the distance between qubits.

\begin{figure}[t]
  \centering
  \includegraphics[width=\columnwidth]{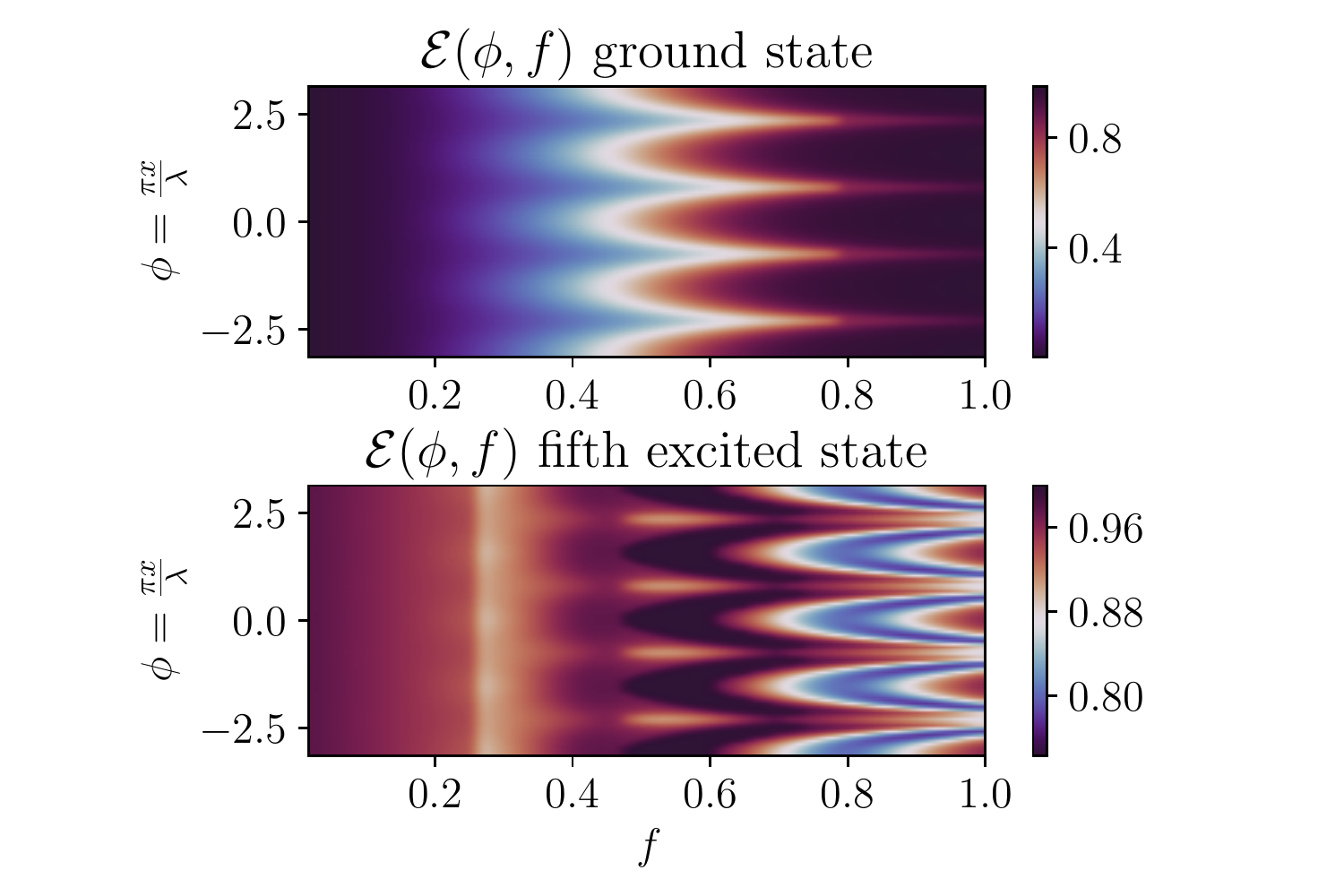}
  \caption{(Color online) The surface of the entanglement
    $\mathcal{E}(\phi, f)$ of the ground state and the fifth excited
    state as a function of the dimensionless coupling constant $f$ and
    the coordinate $\phi$. The parameter $\delta = 1$.}\label{fig:4}
\end{figure}

In order to compute entanglement $\mathcal{E}$ we use the
definition \cite{Entanglement_2006}
\begin{align}
  \mathcal{E}(x,f)
  &= -\sum_{s_{1}} p_{s_{1}}
  \log_{2}p_{s_{1}}, \label{eq:27}
  \\
  p_{s_{1}}
  &= \sum_{n,s_{2}} |C^{0}_{n,s_{1},s_{2}}|^{2}, \label{eq:28}
\end{align}
which is demonstrated in Fig.~\ref{fig:4}. As was expected, for large
values of the coupling constant $f$ the states of both qubits are
entangled and the degree of entanglement depends on the distance
between qubits. The control of the entanglement of different quantum
states by the variation of the distance between qubits can be used for
the encoding of quantum information by two twol-level emitters
\cite{PhysRevA.101.032335}. In addition, an analytical approximation
(\ref{eq:34}) demonstrates that the spin part of the system is
entangled with the field part.


\section{Conclusion}
\label{sec:conclusion}

In our work we have investigated a system of two qubits, which
interact with the quantum electromagnetic field inside a cavity. The
qubits are located on the distances of the order of the wavelength of
the radiation that is sufficiently larger than the corresponding qubit
size. This allowed us to describe the individual qubit within the
dipole approximation. In addition, we have employed the adiabatic
approximation and considered the operator of the kinetic energy of the
qubits as a perturbation. As a consequence, we have approximately
separated the variables in the center of mass reference system of
qubits. In a full analogy with molecular physics we have computed the
radiation induced potential (terms) of the system of two qubits. As a
result, the observable characteristics of the system became functions
of the distance between qubits and the coupling constant $f$ of the
qubits-field interaction.

We have identified the most interesting range of the coupling constant
(USC regime) when the interaction between qubits and the field is the
strongest. We have calculated the observable characteristics
numerically and constructed an analytical approximation. Moreover, we
have demonstrated that by varying the distance between qubits the
observable characteristics are strongly changing. In addition, by
changing the qubits positions one can perform a quantum control of the
system, which is useful for the recording and the transmission of
quantum information. For example, this can be important in the
following applications: scattering of the resonant radiation
\cite{Scattering_2020}; dynamics of an
entanglement of two-level emitters \cite{PhysRevA.101.032335}; to
control characteristics of spontaneous emission in two level systems
\cite{feranchuk_spontaneous_2017}; appearance of the periodic
structure in the interaction potential in the system of $N$ atoms ---
the continuous Dicke model \cite{PhysRevA.99.013839}, which should be
investigated beyond the rotating wave approximation.

\appendix

\section{Analytical approximation for energy levels}
\label{appendix:A}

In this appendix we will derive an analytical approximation for
eigenvalues and eigenvectors of the system of two two-level atoms
interacting with a single mode quantum field.

Let us first consider the case when $\delta$ equals zero in the
Hamiltonian of the system Eq.~(\ref{eq:14}). In this case the problem
becomes exactly solvable, since both spins are diagonalized by one of
the following wave functions $\chi_{\uparrow}^{1}\chi_{\uparrow}^{2}$,
$\chi_{\uparrow}^{1}\chi_{\downarrow}^{2}$,
$\chi_{\downarrow}^{1}\chi_{\uparrow}^{2}$,
$\chi_{\downarrow}^{1}\chi_{\downarrow}^{2}$, where $\chi_{\uparrow}$,
$\chi_{\downarrow}$ are eigenvectors of $\sigma_{1}$. The field part
of the Hamiltonian contains only the first powers of operators. As is
well known, in this case the field Hamiltonian is diagonalized by
displacing the classical component from the field operators or in
other words by performing a unitary transformation of the Hamiltonian
with the operator of the coherent state
$\hat{D}(u) = \exp(u\hat{a}^{\dag} - u^{*}\hat{a})$ that transforms
the field operators as
\begin{equation}
  \begin{aligned}
    \hat{D}^{\dag}(u) \hat{a} \hat{D}(u)
    &= \hat{a} + u,
    \\
    \hat{D}^{\dag}(u) \hat{a}^{\dag} \hat{D}(u)
    &= \hat{a}^{\dag} + u^{*}.
  \end{aligned} \label{eq:29}
\end{equation}

The parameter $u$ is then chosen from the condition that the first
power of the operators vanishes. By performing the unitary
transformation of the field part of the Hamiltonian we can find out
the following expressions for each of the four spin wave functions
listed above
\begin{align}
  \hat{H}_{1\mathrm{f}}
  &= \hat{a}^{\dag}\hat{a} + u_{1}(\hat{a} + \hat{a}^{\dag}) + u_{1}^{2}
    \nonumber
  \\
  &+ 2f(\hat{a} + \hat{a}^{\dag})\cos{\phi} + 4u_{1}f
    \cos{\phi}, \label{eq:30}
  \\
  \hat{H}_{2\mathrm{f}}
  &= \hat{a}^{\dag}\hat{a} - iu_{2}(\hat{a} -
    \hat{a}^{\dag}) + u_{2}^{2} \nonumber
  \\
  &+ 2if(\hat{a} - \hat{a}^{\dag})\sin{\phi} -
    4u_{2}f \sin{\phi}, \label{eq:31}
\end{align}
\begin{align}
  \hat{H}_{3\mathrm{f}}
  &= \hat{a}^{\dag}\hat{a} - iu_{3}(\hat{a} - \hat{a}^{\dag}) + u_{3}^{2}
    \nonumber
  \\
  &- 2if(\hat{a} - \hat{a}^{\dag})\sin{\phi} +
    4u_{3}f \sin{\phi}, \label{eq:32}
  \\
  \hat{H}_{4\mathrm{f}}
  &= \hat{a}^{\dag}\hat{a} + u_{4}(\hat{a} + \hat{a}^{\dag}) +
    u_{4}^{2} \nonumber
  \\
  &- 2f(\hat{a} + \hat{a}^{\dag})\cos{\phi} -
    4u_{4}f \cos{\phi}. \label{eq:33}
\end{align}
Here we considered that parameters of the coherent states for the
cases $1,4$ are pure real and for the cases $2,3$ are pure imaginary
numbers. From here we can conclude that $u_{1,4} = \pm 2f \cos{\phi}$
and $u_{2,3} = \pm 2if \sin{\phi}$.
\begin{figure}[t]
  \centering
  \includegraphics[width=\columnwidth]{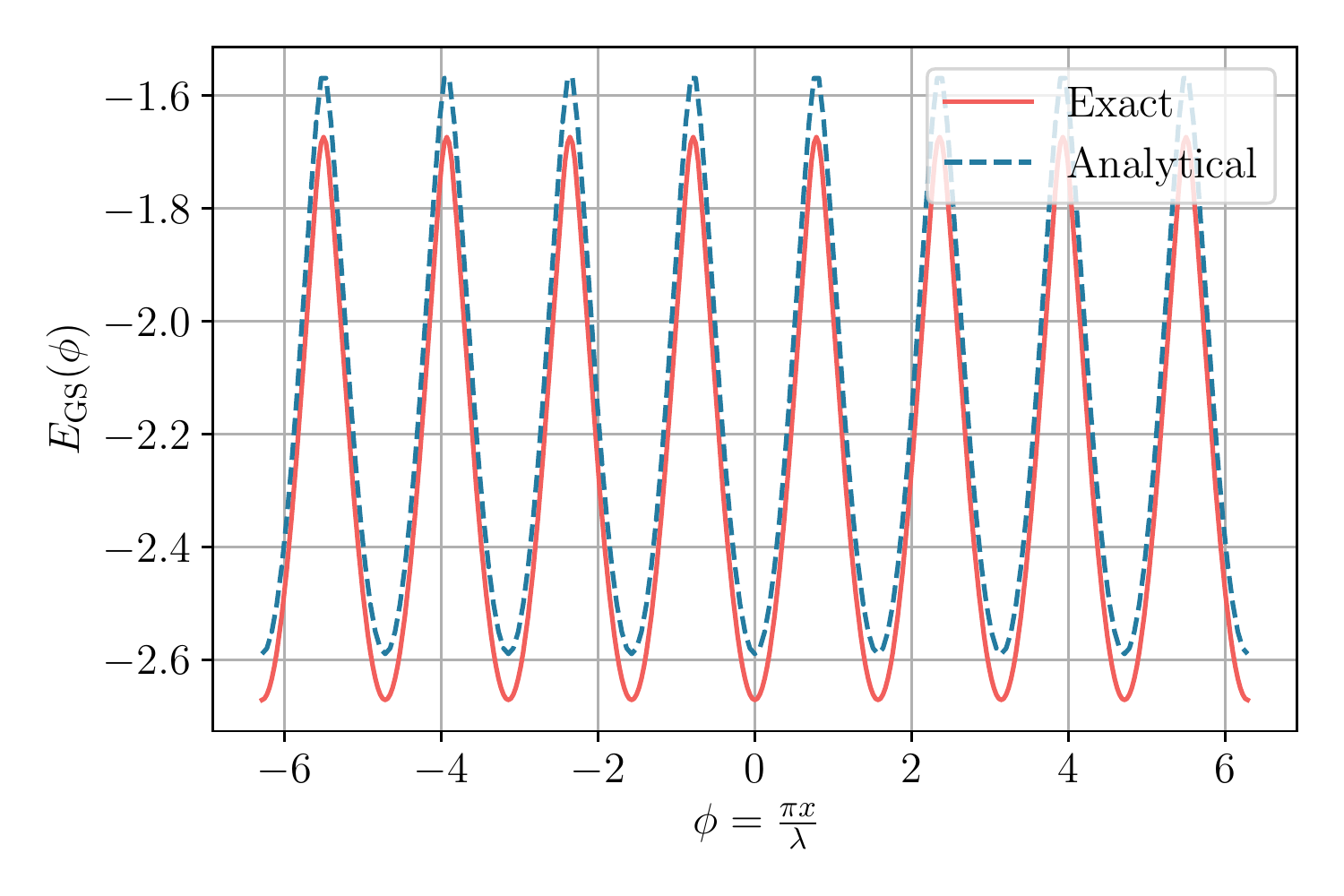}
  \caption{(Color online) The comparison of the exact numerical
    solution with the approximate analytical result of the ground
    state energy $E_{\mathrm{GS}}$ as a function of the distance
    $\phi$ between two atoms. The parameter $\delta = 1$ and the USC
    regime is considered for $f = 0.8$.}\label{fig:5}
\end{figure}

Now let us consider the general situation of $\delta \neq 0$. In this
case the spins and the field are entangled and the approximate wave
function of the system is expressed as a linear combination of the
previously found four possibilities, namely
\begin{align}
  |\Psi\rangle
  &= A_{1} \chi_{\uparrow}^{1}\chi_{\uparrow}^{2}
    |n, 2f\cos{\phi}\rangle \nonumber
  \\
  &+ A_{2} \chi_{\uparrow}^{1}\chi_{\downarrow}^{2} |n, -2if
    \sin{\phi}\rangle  \nonumber
  \\
  &+ A_{3}\chi_{\downarrow}^{1}\chi_{\uparrow}^{2} |n, 2if
    \sin{\phi}\rangle \nonumber
  \\
  &+ A_{4} \chi_{\downarrow}^{1}\chi_{\downarrow}^{2}
  |n, -2f\cos{\phi}\rangle, \label{eq:34}
\end{align}
where we introduced the notation $|n, u\rangle =
\hat{D}(u)|n\rangle$. As a result, the energy of the system is given as
the solution of the eigenvalue problem
$\hat{H} |\Psi\rangle = E|\Psi\rangle$, with the Hamiltonian $\hat{H}$
defined by Eq.~(\ref{eq:14}) in the finite basis, consistent of four
wave functions. The solution of this eigenvalue problem leads to the
desired coefficients $A_{i}$, $i = 1..4$ and the eigenvalues. For
example, for the first four states one finds the following expressions
for the Hamiltonian matrix (here we employed the notation for the
coherent states $|0, u\rangle = |u\rangle$)
\begin{widetext}
  \begin{align}
    H_{ij} = \begin{pmatrix}
      u_{1}^{2} + 4f u_{1} \cos{\phi} & \frac{\delta}{2}\langle
      u_1|iu_2\rangle & \frac{\delta}{2} \langle u_1|iu_3\rangle & 0
      \\
      \frac{\delta}{2}\langle i u_2|u_1\rangle & (u_2^2 - 4 f
      u_2\sin\phi) & 0 & \frac{\delta}{2} \langle i u_2|u_4\rangle
      \\
      \frac{\delta}{2} \langle i u_3|u_1\rangle & 0 & (u_3^2 + 4 f
      u_3\sin\phi) & \frac{\delta}{2} \langle{i u_3|u_4}\rangle
      \\
      0 & \frac{\delta}{2}\langle{u_4|iu_2}\rangle &
      \frac{\delta}{2}\langle{u_4|iu_3}\rangle & (u_4^2 - 4 f u_4\cos
      \phi)
    \end{pmatrix}, \label{eq:35}
  \end{align}
\end{widetext}
where the overlapping integrals between different coherent states are
defined as
\begin{align*}
  \langle{v|u}\rangle
  &= \langle 0| e^{v^* \hat{a} - v \hat{a}^{\dag}}
    e^{u \hat{a}^{\dag} - u^* \hat{a}}|0 \rangle
  \\
  &= \langle 0|e^{(u- v)\hat{a}^{\dag} - (u^* - v^*)\hat{a}} |0\rangle
    e^{1/2 (u v^* - v u^*)}
  \\
  &= e^{-1/2 |u - v|^2} e^{1/2 (u v^* - v u^*)}.
\end{align*}

By diagonalizing the matrix (\ref{eq:35}) of the system one finds the
following expressions for the energy levels of the first four states
\begin{align*}
  E_1
  &= - 2 f^2 - \sqrt{4f^4\cos^2 2\phi + \delta^2 e^{- 4f^2}},
  \\
  E_2
  &= - 4 f^2 \sin^2\phi,
  \\
  E_3
  &= - 4 f^2 \cos^2\phi,
  \\
  E_4
  &= - 2 f^2 + \sqrt{4f^4\cos^2 2\phi + \delta^2 e^{- 4 f^2}}.
\end{align*}

Analogous formulas can be obtained for other states of the system.

Finally, we pay attention to the fact that the period of oscillations
is equal to $\pi / 2$, despite that fact that the Hamiltonian of the
system has periodicity $2\pi$ in the dimensionless distance $\phi$
between qubits. In Fig.~\ref{fig:5} we plot the slice of the potential
surface for $f = 0.8$ and compare the numerical versus analytical
energy of the ground state. As can be concluded from the figure the
analytical approximation describes well all the qualitative properties
of the system and agrees with the exact numerical solution.

\bibliography{biblnrwa}

\end{document}